\documentclass[aps,pre,showpacs,twocolumn]{revtex4-1}

\usepackage{graphicx}
\usepackage{bm}

\begin{document}

\title{First passage time for subdiffusion: The nonadditive entropy approach\\ versus the fractional model}

\author{Tadeusz Koszto{\l}owicz}
  \email{tadeusz.kosztolowicz@ujk.kielce.pl}
 \affiliation{Institute of Physics, Jan Kochanowski University,\\
         ul. \'Swi\c{e}tokrzyska 15, 25-406 Kielce, Poland.}

\author{Katarzyna D. Lewandowska}
  \email{kale@gumed.edu.pl}
 \affiliation{Department of Radiological Informatics and Statistics, Medical University of
         Gda\'nsk,\\ ul. Tuwima 15, 80-210 Gda\'nsk, Poland.}

\begin{abstract}
We study the similarities and differences between different models concerning subdiffusion. More particularly, we calculate first passage time (FPT) distributions for subdiffusion, derived from Greens' functions of  nonlinear equations obtained from Sharma--Mittal's, Tsallis's and Gauss's nonadditive entropies. Then we compare these with FPT distributions calculated from a fractional model using a subdiffusion equation with a fractional time derivative. All of Greens' functions give us exactly the same standard relation $\left\langle (\Delta x)^2\right\rangle =2 D_\alpha t^\alpha$ which characterizes subdiffusion ($0<\alpha<1$), but generally FPT's are not equivalent to one another. We will show here that the FPT distribution for the fractional model is asymptotically equal to the Sharma--Mittal model over the long time limit only if in the latter case one of the three parameters describing Sharma--Mittal entropy $r$ depends on $\alpha$, and satisfies the specific equation derived in this paper, whereas the other two models mentioned above give different FTPs with the fractional model. Greens' functions obtained from the Sharma--Mittal and fractional models -- for $r$ obtained from this particular equation --  are very similar to each other. We will also discuss the interpretation of subdiffusion models based on nonadditive entropies and the possibilities of experimental measurement of subdiffusion models parameters.
\end{abstract}


\maketitle

\section{Introduction}

Over about the last 20 years the anomalous diffusion process has been observed in many physical systems. Simultaneously, various theoretical models of this process have been put forward (see \cite{bg,mk,mk1} and the references cited therein). It is worth considering what the issue of anomalous diffusion is and what definition of this process can be taken into account. We note that the situation is different  here compared to the normal diffusion process. Namely, in the latter case the different models produce results which are equivalent to each other. For example, the stochastic approach provides the same normal diffusion equation as entropy formalism, and their fundamental solutions appear to be Gaussian functions with their second moment proportional to time 
	\begin{equation}\label{eq1_a}
\left\langle (\Delta x)^2\right\rangle =2Dt\;.
	\end{equation}
Consequently, the question: `What is the definition of normal diffusion?' can be answered in a few equivalent ways, such as with the process described by the Langevin equation with white noise, the standard Wiener process, the random walk of a particle where the probability distributions of its step length and the waiting time to take its next step have finite moments, or the process described by a probability density which maximizes the Boltzmann--Shannon entropy. Although from a mathematical point of view these definitions are not exactly equivalent to each other, physicists usually treat these definitions equivalently. Anomalous diffusion is a process which qualitatively differs from normal diffusion, so all of the above mentioned definitions are not fulfilled. However, such treatment causes ambiguity in the definitons.
There arises a problem with the definition of anomalous diffusion as a process which denies normal diffusion definition.  
The Wiener process is replaced by the frational one (consequently the Langevin equation is changed), whithin the Continuous Time Random Walk (CTRW) formalism at least one of the probability distributions describing a single particle jump has infinite moments, and nonadditive entropies are used instead of Boltzmann--Shannon entropy. 
The anomalous diffusion is frequently defined by its interpretation as a non--Markovian random walk which is described by non--Gaussian probability distribution. However, this definition is limited to the stochastic processes and its relation to thermodynamics or deterministic models is not obvious. To find a more general definition one sholud assume that anomalous diffusion is characterized by a special parameter (in the following denoted by $\alpha$) which is a `measure' of how far the anomalous diffusion process is from the normal diffusion one. Since within the CTRW formalism the anomalous diffusion model provides us with the probability distributions with the second moment to be nonlinear of time
	\begin{equation}\label{eq1}
\left\langle (\Delta x)^2\right\rangle =2D_\alpha t^\alpha\;,
	\end{equation}
where $\alpha>0,\; \alpha\ne1$, the statement that the model which provide the relation (\ref{eq1}) {\it can be treated} as an anomalous diffusion model has been rather widely used. The parameter $\alpha$ plays a different role in the models; it is related to the fractional derivative order in the anomalous diffusion equations or it controls a measure of nonadditivity of the entropy.

The most used anomalous diffusion models seems to be the CTRW model and the models derived from nonadditive entropies. Throughout this paper we will refer to the model based on CTRW as a fractional model. CTRW provides the linear anomalous diffusion equation with the fractional-order derivatives \cite{mk,mk1,compte}, while the models based on the nonadditive entropies give nonlinear differential (or integral--differential) equations with derivatives of a natural order \cite{frank,tsallis,pp,bppp,cj,tb,dwt,frank3,sw,fd}.  The simplest stochastic interpretation of the anomalous diffusion seems to be found within CTRW models, where the random walker waits an anomalously long time to make its next step, for example in the transport process of sugars in gel (for subdiffusion) or where the step length can be anomalously long with a relatively high probability, for example a random walk in a turbulent medium (for superdiffusion). For the nonadditive entropy model, a physical meaning of the anomalous diffusion equation is manifested mainly in its stationary version, namely, the stationary solution of the equation maximizes nonadditive entropy. 
However, as is shown in \cite{dgn}, there is a non--Markovian process which provides the relation (\ref{eq1_a}). Thus, the stochastic definition of anomalous diffusion is, in some situations, in contradiction with the ones based on the relation (\ref{eq1}).

We note that in a system where the relation (\ref{eq1}) is valid, other functions, which can be measured experimentally, of a type $f(t)\sim t^{\alpha/2}$ describe the system, such as the time evolution of near membrane layer thickness \cite{kdm,kdm1} and the time evolution of the reaction front in the subdiffusive system with chemical reactions \cite{yal,kl1}. Let us also note that there are models which do not have fully satisfactory stochastic interpretations yet, as, for example the anomalous diffusion process described within nonadditive entropy formalism; but in these models we can also find the power--like important characteristic of the system identical with the ones found within stochastic models. The stochastic models are based on the assumptions which simplify the problem which the experimental data is not capable of confirming,
as the assumptions concerning random walk of a particle for a separable case of CTRW formalism. On the other hand, the stochastic interpretation of various diffusion models is needed. The attempts to find stochastic interpretation of nonadditive entropy formalism have mainly been made using the modified Langevin equation for the description of the anomalous diffusion process which is simultaneously described by a nonlinear differential equation. The result is that a random force occurring in the Langevin equation depends on the solution to the nonlinear equation \cite{wt,frank2,frank4,borland,stariolo}. This situation can be interpreted as the existence of feedback between a system and a random force, whose interpretation is --- at least in our opinion --- not obvious. We remark here that such nonlinear equations were also derived from the master equation \cite{cn}. Recently, we found a new stochastic interpretation of subdiffusion as a `long memory diffusion' described by the generazlied linear Langevin equation in a system with external Gamma--type noise \cite{kl2012}. The anomalous diffusion equation derived within this new model perfectly coincides with the one derived within the nonadditive Sharma--Mittal entropy formalism.

The above considerations lead us to take the following definition of the anomalous diffusion: {\it An equation whose fundamental solution (Green's function) $G(x,t;x_0)$ provides the relation (\ref{eq1}) where
	\begin{displaymath}
\left\langle (\Delta x)^2\right\rangle =\int_{-\infty}^{\infty}(x-x_0)^2G(x,t;x_0)dx\;,
	\end{displaymath}
can be treated as an equation describing the anomalous diffusion process}. Throughout this paper we have assumed that we are dealing with one-dimensional homogeneous systems without any external fields and convective flows; here $D_\alpha$ is the anomalous diffusion coefficient measured in the units $m^2/s^\alpha$ and $\alpha$ is the anomalous diffusion parameter; $0<\alpha<1$ for subdiffusion, $\alpha>1$ for superdiffusion. 
Thus, anomalous diffusion is controlled by two parameters: $D_{\alpha}$ and $\alpha$. We would like to add that the anomalous diffusion coefficient is sometimes defined differently, for example the subdiffusive coefficient $\tilde{D}_\alpha$ is often defined as \cite{mk,mk1}
$\tilde{D_\alpha}= \Gamma(1+\alpha)D_\alpha$.	
Although Greens' functions derived from the equation mentioned above provide Eq. (\ref{eq1}), other important characteristics, such as a first passage time distribution, are different. 

First passage time (FPT), which is one of the most important characteristics in normal and anomalous diffusions \cite{render,hughes,klckm}, was studied for anomalous diffusion mainly within the fractional model \cite{mk,mk1,barkai,mkbj,g,yl,lab,dgh,cmgkt,kck,akk,lels,fl,kl} as well as in the lattice and fractal medium \cite{ctvbk,cbk}. The FPT is defined as the time that the random walker takes to reach a target located in $x_M$ for the first time, from the starting point $x_0$. The FPT has been used to describe real physical processes such as animals searching for food \cite{bcmsv}, the passage of DNA molecules through a membrane channel \cite{mkbj} and the spreading of disease \cite{l} etc. Moreover, the FPT distribution can be used to calculate other characteristic functions which are measured experimentally, such as the time evolution of an amount of a substance leaving a sample or the substance flux flowing through a sample surface.

In this paper we study subdiffusive systems. We derive the FPT distributions for subdiffusion equations derived form Sharma-Mittal, Tsallis and Gauss nonadditive entropies and compare them with the result obtained from the fractional model. We sholud add here that we will pay particular attention to the nonlinear differential anomalous diffusion equations derived from nonadditive entropies and we will not discuss the interpretation of the entropies. Our strategy is as follows: we adapt Greens' functions presented in their generalized forms in Frank's book  \cite{frank} into special forms so that each of them exactly satisfies the relation (\ref{eq1}). Next, we calculate the FPT distribution functions and compare them with the one obtained from the fractional model. We derive consistency conditions in the functions calculated from various models over the long time limit. We briefly discuss the similarity of Greens' functions and the interpretation of the parameters occurring in nonadditive entropies and their connection with the parameters $D_\alpha$ and $\alpha$, which are measured experimentally. This method would allow us to measure nonadditive entropy parameters. 

\section{The method}

Here we will present the functions and equations which define Green's function and distribution of first passage time used in our considerations. The description of nonlinear anomalous diffusion equations derived from nonadditive entropies and their analytical solutions (Greens' functions) are based on Frank's book \cite{frank}.

\subsection{Anomalous diffusion equations}

\subsubsection{Nonadditive entropy formalism}

The Sharma--Mittal entropy is defined as
	\begin{equation}\label{smen}
S_{q,r}^{SM}[P]=\frac{1-\left(\int P^r dx\right)^{(q-1)/(r-1)}}{q-1}\;,
	\end{equation}
where $q,r>0$, $q,r\neq 1$, $P$ denotes a probability of finding a particle at the point $x$ at time $t$.
Gauss entropy obtained in the limit $r\rightarrow 1^-$ reads
	\begin{displaymath}
S^G_q[P]=\frac{1-{\rm e}^{(q-1)\int{P{\rm ln}Pdx}}}{q-1}\;,
	\end{displaymath}
where $q>0$, $q\neq 1$. 
The Tsallis entropy can be obtained from (\ref{smen}) putting $q=r$.
For two statistically independent systems $A$ and $B$ the Sharma--Mittal entropy satisfies the following equation
	\begin{eqnarray*}
S_{q,r}^{SM}(A+B)&=&S_{q,r}^{SM}(A)+S_{q,r}^{SM}(B)\\&+&(1-q)S_{q,r}^{SM}(A)S_{q,r}^{SM}(B)\;.
	\end{eqnarray*}
For $q\neq 1$ one deals with nonadditive entropy \cite{tsallisentropy,tt}.

There are two main ways to derive anomalous diffusion equation from nonadditive entropy. Within the first one the stationary state is generated by means of a maximum entropy condition under conditions which assume that the $q$-moments are finite.
Within the second method the flux $J$ defined by the equation $J=L(P)(\delta S/\delta P)$ is combined with the continuity equation $\partial C/\partial t=-\partial J/\partial x$ \cite{cj}. For the normal diffusion case $L(P)\sim 1/P$. There is no obvious choice of the function $L(P)$ for anomalous diffusion.
The anomalous diffusion equation derived within nonadditive entropy formalism reads \cite{frank}
	\begin{equation}\label{smeq}
\frac{\partial P(x,t)}{\partial t}=Q_i\Psi_i[P]\frac{\partial^2 P^r(x,t)}{\partial^2 x}\;,
	\end{equation}
where $r\neq 1$ and $r>1/3$, $r=q$ for the Tsallis case and $r=1$ for the Gauss case, the index $i$ denotes the symbol identifying entropy, $Q_i$ denotes the fluctuation strength,
and	
	\begin{displaymath}
\Psi_{SM}[P]=\left(\int{P^rdx}\right)^{\frac{q-r}{r-1}},\qquad\Psi_T[P]=1\;,
	\end{displaymath}
	\begin{displaymath}
\Psi_G[P]={\rm e} ^{(q-1)\int{P{\rm ln} P dx}}\;.
        \end{displaymath}

\subsubsection{Fractional model}

Within the separable case of CTRW formalism it is assumed that a particle takes its next step of a length $\rho$ after time $\tau$, where both are independent random variables. For subdiffusion the probability distribution $\omega(\tau)$ is of `thick tail', $\omega(\tau)\approx -\tau_\alpha /t^{1+\alpha}\Gamma(-\alpha)$ for a sufficienty large time (the mean value of $\omega$ equals infinity), whereas $\lambda(\rho)$ is the Gaussian distribution, $\lambda(\rho)=\exp\left({-\rho^2/2\sigma^2}\right)/\sqrt{2\pi\sigma^2}$. Under the above assumptions one obtains the linear subdiffusion differential equation with the Riemann-Liouville fractional derivative \cite{mk,mk1,compte}
	\begin{equation}\label{eq10}
\frac{\partial P(x,t)}{\partial t}=\tilde{D}_\alpha\frac{\partial^\alpha}{\partial t^\alpha}\frac{\partial^2 P(x,t)}{\partial x^2}\;,
	\end{equation}
where $\tilde{D}_\alpha=\sigma^2/\tau_\alpha$, the Riemann--Liouville fractional derivadive is defined as follows for $\alpha>0$
	\begin{displaymath}
\frac{d^\alpha f(t)}{dt^\alpha}=\frac{1}{\Gamma(n-\alpha)}\frac{d^n}{dt^n}\int_0^t{(t-t')^{n-\alpha-1}f(t')dt'}\;,
	\end{displaymath}
where $n$ is a natural number fulfilled $\alpha\leq n<\alpha+1$.

\subsection{Green's function}

Green's function (GF) is defined here as a solution to the appropirate diffusion equation with the initial condition
	\begin{displaymath}
G(x,0;x_0)=\delta(x-x_0)\;,
	\end{displaymath}
$\delta$ denotes the delta-Dirac function. When particles are independently transported and all of them start their movement 
at $x_0$ at the initial moment $t=0$, Green's function can be interpreted as a concentration profile of the particles normalized to one (i.e. divided by the number of particles). Another interpretation of the GF is that it is treated as a probability density of finding a particle at point $x$ at time $t$ under the condition that its initial position is $x_0$. We should add here, that there is a stochastic interpretation of subdiffusive movement of a particle described by a nonlinear differential equation, which assumes that the particle is transported independently of other particles \cite{kl2012}. For the unrestricted system the GF satisfies natural boundary conditions which require the disappearance of the function at an infinite distance from the initial position
$G(x,t;x_0)\rightarrow 0$,
when $x\rightarrow\pm\infty$. 

Green's function for the nonlinear diffusion equation ($m>0$, $m\neq 1$)
	\begin{displaymath}
\frac{\partial P(x,t)}{\partial t}=\frac{\partial^2 P^m(x,t)}{\partial x^2}\;,
	\end{displaymath}
known as the Barenblatt solution, reads \cite{wu}
	\begin{equation}\label{bar}
G(x,t;x_0)=t^{-k}\left[\left\{1-\frac{k(m-1)|x-x_0|^2}{2mt^{2k}}\right\}_+\right]^{\frac{1}{m-1}}\;,
	\end{equation}
where $k=1/(m+1)$, $\{u\}_+={\rm max}\{u,0\}$.

In our paper we use Green's function for the system with a fully absorbing wall.
The commonly used boundary condition at an absorbing wall reads
	\begin{displaymath}
G_{\rm abs}(x_M,t;x_0)=0\;.
	\end{displaymath}
Due to the interpretation of GF and the symmetry arguments, the GF for the normal and subdiffusive systems with a fully absorbing wall can be found through the means of the method of images, which for $x,x_0<x_M$ gives
	\begin{equation}\label{eq6}
G_{\rm abs}(x,t;x_0)=G(x,t;x_0)-G(x,t;2x_M-x_0)\;.
	\end{equation}

\subsubsection{Nonadditive entropy formalism}

Greens' functions of (\ref{smeq}) for the Sharma--Mittal and Tsallis models take the form of Eq. (\ref{bar}) and read \cite{frank}
	\begin{eqnarray}\label{smgf}
\lefteqn{G_{SM}(x,t;x_0)=}\nonumber\\
&=&D_{SM}(t)\left[\left\{1-\frac{C_{SM}(t)}{2}(r-1)(x-x_0)^2\right\}_+\right]^{\frac{1}{r-1}}\;,\nonumber\\
&&
	\end{eqnarray}
where $r>1/3$, $r,q \neq 1$, $q>0$,
	\begin{eqnarray}\label{tgf}
\lefteqn{G_T(x,t;x_0)=}\nonumber\\
&=&D_T(t)\left[\left\{1-\frac{C_T(t)}{2}(q-1)(x-x_0)^2\right\}_+\right]^{\frac{1}{q-1}}\;,\nonumber\\
&&
	\end{eqnarray}
where $q>1/3$, $q \neq 1$,
	\begin{equation}\label{ggf}
G_G(x,t;x_0)=D_G(t){\rm exp}\left(-\frac{C_G(t)}{2}(x-x_0)^2\right)\;.
	\end{equation}
The functions occurring in (\ref{smgf})--(\ref{ggf}) are defined as
	\begin{equation}\label{smd}
D_{SM}(t)=\left[\frac{1}{2r(1+q)Q_{SM}K_{r,q}\left|z_r\right|^2t}\right]^{\frac{1}{1+q}}\;,
	\end{equation}
	\begin{equation}\label{td}
D_T(t)=\left[\frac{1}{2q(1+q)Q_T\left|z_q\right|^2t}\right]^{\frac{1}{1+q}}\;,
        \end{equation}
	\begin{equation}\label{td1}
D_G(t)=\left[\frac{{\rm e}^{(q-1)/2}}{2\pi(1+q)Q_G t}\right]^{\frac{1}{1+q}}\;,
        \end{equation}
where
	\begin{equation}\label{eq14}
z_r=\left\{
\begin{array}{ll}
\sqrt{\frac{\pi}{r-1}}\frac{\Gamma(r/(r-1))}{\Gamma((3r-1)/(2(r-1)))}\;, & r>1\;,\\
\sqrt{\pi}\;, & r=1\;,\\
\sqrt{\frac{\pi}{1-r}}\frac{\Gamma((1+r)/2(1-r))}{\Gamma(1/(1-r))}\;, & 1/3<r<1\;.
\end{array}
\right.
	\end{equation}
and
	\begin{equation}\label{defK}
K_{r,q}=\left\{
\begin{array}{ll}
\left(\frac{3r-1}{2r}\right)^{\frac{q-r}{1-r}}\;, & r\ne1\;,\\
{\rm e}^{(1-q)/2}\;, & r=1\;,
\end{array}
\right.
	\end{equation}		
	
	\begin{eqnarray}\label{c}
C_{SM}(t)&=&2(z_r D_{SM}(t))^2\;,\nonumber\\
 C_T(t)&=&2(z_q D_T(t))^2\;,\\
 C_G(t)&=&2\pi(D_G(t))^2\;.\nonumber
	\end{eqnarray}	
The second moment of Greens' functions presented above can be calculated according to the formulae
	\begin{eqnarray}\label{smk}
\left\langle\left(\Delta x\right)^2(t)\right\rangle_{SM}&=&\frac{2}{3r-1}\frac{1}{C_{SM}(t)},\nonumber\\
\left\langle\left(\Delta x\right)^2(t)\right\rangle_{T}&=&\frac{2}{3q-1}\frac{1}{C_{T}(t)},\\
\left\langle\left(\Delta x\right)^2(t)\right\rangle_{G}&=&\frac{1}{C_{G}(t)}.\nonumber
	\end{eqnarray}

\subsubsection{Fractional model}

For Eq. (\ref{eq10}) the Green's function is (in the following, the index $F$ corresponds to the fractional model)
	\begin{equation}\label{eq34}
G_F(x,t;x_0)=\frac{1}{2\sqrt{\tilde{D}_\alpha}}f_{\alpha/2-1,\alpha/2}\left(t;\frac{|x-x_0|}{\sqrt{\tilde{D_\alpha}}}\right)\;,
	\end{equation}
where 
	\begin{equation}\label{f}
f_{\nu,\beta}(t;a)=\frac{1}{t^{1+\nu}}\sum_{k=0}^\infty\frac{1}{\Gamma(-k\beta-\nu)k!}\left(-\frac{a}{t^\beta}\right)^k\;,
        \end{equation}
$\beta,a>0$, the function $f$ can also be expressed in terms of the Fox function \cite{tk}. We should add here, that the methods of solving Eq.~(\ref{eq10}) are presented, among others, in \cite{mk,kdm,tk,p}.

\subsection{First passage time}

Let us assume that a particle is located at $x_0$ at the initial moment $t=0$. The time when the particle reaches the point $x_M$ for the first time is a random variable described by a probability density of $F$, calculated according to the formula for $t>0$
	\begin{equation}\label{eq7}
F(t;x_0,x_M)=-\frac{dR(t;x_0,x_M)}{dt}\;,
	\end{equation}
for $t\le0$ we put $F(t;x_0,x_M)=0$, and where $R(t;x_0,x_M)$ denotes the probability of finding the particle at time $t$  starting from $x_0$ in the system with a fully absorbing wall located at $x_M$  (in the following we assume that $x_0<x_M$)
	\begin{equation}\label{eq8}
R(t;x_0,x_M)=\int_{-\infty}^{x_M}{G_{\rm abs}(x,t;x_0)dx}\;.
	\end{equation}
The cumulative function of $F$ is $1-R(t;x_0,x_M)$.
The mean first time $\left\langle T\right\rangle$ is defined as
	\begin{displaymath}
\left\langle T\right\rangle=\int_0^\infty t F(t;x_0,x_M)dt\;.
	\end{displaymath}
Assuming that $tR(t;x_0,x_M)\rightarrow 0$ when $t\rightarrow\infty$, one obtains
	\begin{equation}\label{eq11}
\left\langle T\right\rangle=\int_0^\infty R(t;x_0,x_M)dt\;.
	\end{equation}

\section{Greens' functions generated relation (\ref{eq1}) and first passage time distributions}

Greens' functions (\ref{smgf}), (\ref{tgf}) and (\ref{ggf}), presented in the previous section, depend on the parameters $q$ (which can be interpreted as a measure of entropy nonadditivity) and the fluctuation strength $Q_i$, here the index $i$ denotes the model, $i=SM,T,G$ for the Sharma--Mittal, Tsallis and Gauss models, respectively. Sharma--Mittal Green's function additionally depends on the parameter $r$. Green's function for the Tsallis and Gauss models can be treated as specific cases of the Sharma--Mittal one, using $q=r$ for the Tsallis model and having as a limit $r\rightarrow 1^-$ for the Gauss model. However, retaining the commonly used terminology, we consider these functions separately. 

We look at two sets of subdiffusive models having different physical origins. The first contains the models derived from nonadditive entropy and the second contains the fractional model. In general, both of these sets can describe the same physical processes. However, processes also occur which can be described by models from one set alone, the other set of models cannot be applicable in describing such a process (this problem will be briefly discussesd in the Final Remarks). We are going to find the accordance conditions between models from both sets using the FPT distributions.
Our further considerations are based on the following assumptions: the first, since the models describe the same subdiffusion process, all of Greens' functions should provide the same relation which defines subdiffusion (\ref{eq1}); the second, the parameters $\alpha$ and $D_\alpha$ are measured experimentally. The examples of such measurements, where the empirical results were compared with theoretical functions derived within the fractional model, are presented in \cite{kdm,kdm1,kmk}. 
Taking into account Eqs. (\ref{eq1}), (\ref{smd}), (\ref{td}), (\ref{td1}), (\ref{c}), and (\ref{smk}) we come to the conclusion that relation (\ref{eq1}) will be satisfied by all of Greens' functions $G_{i}(x,t;x_0)$ when 
	\begin{equation}\label{defq}
q=\frac{2}{\alpha}-1
	\end{equation}
for all models, while the fluctuation strength must be chosen for each model separately. The fluctuation strength $Q_i$, which in general depends on $q$ and $r$, plays the key role in expressing the parameters of the models based on nonadditive entopy by  $\alpha$ and $D_\alpha$.
When $q=r=1$ we are dealing with normal diffusion and then $Q_i$ is identified as the normal diffusion coefficient, $D_1=Q_i$ for each model. Taking into account (\ref{defq}) they read
	\begin{equation}\label{defQi}
Q_{SM}=\frac{\alpha\left[2D_\alpha(3r-1)\right]^{1/\alpha}}{4rK_{r,2/\alpha-1}|z_r|^{2(1-1/\alpha)}}\;,
        \end{equation}
        \begin{equation}\label{defQT}
Q_{T}=\frac{\alpha^2\left[2D_\alpha(6/\alpha-4)\right]^{1/\alpha}}{4(2-\alpha)|z_{2/\alpha-1}|^{2(1-1/\alpha)}}\;,
        \end{equation}
        \begin{equation}\label{defQG}
Q_{G}=\frac{\alpha (2D_\alpha)^{1/\alpha}}{2\left(\sqrt{2\pi{\rm e}}\right)^{2(1-1/\alpha)}}\;.
	\end{equation}
In the following subsections, we apply Eqs.~(\ref{defq}) -- (\ref{defQG})  to eliminate the parameters $q$ and $Q_{i}$ from Greens' functions (\ref{smgf}), (\ref{tgf}) and (\ref{ggf}) and then we use Greens' modified functions to calculate the functions $F_i(t;x_0,x_M)$ and $R_i(t;x_0,x_M)$ from (\ref{eq7}) and (\ref{eq8}). In the last subsection we will find these functions for the fractional model. 

\subsection{The Sharma--Mittal model}

Using Eqs.~(\ref{smgf}), (\ref{defq}), and (\ref{defQi}), Green's function for the homogeneous unrestricted system, provided by the Sharma--Mittal entropy model, reads
	\begin{eqnarray}\label{eq12}
\lefteqn{G_{SM}(x,t;x_0)=}\nonumber\\
&=&\frac{1}{\sqrt{2D_\alpha (3r-1)t^\alpha}|z_r|}\left[\left\{1-\frac{(r-1)(x-x_0)^2}{2D_\alpha (3r-1)t^\alpha}\right\}_+\right]^{\frac{1}{r-1}}\;,\nonumber\\
&&
	\end{eqnarray}
where $r>1/3$ and $r\ne1$.
The function has different properties depending on the value of the parameter $r$. In the following, we consider the cases $r>1$ and $1/3<r<1$ separately.

\subsubsection{The case of $r>1$}\label{sec.3.1.1}

\begin{figure}
	\centering
	\includegraphics[angle=-90,scale=0.32]{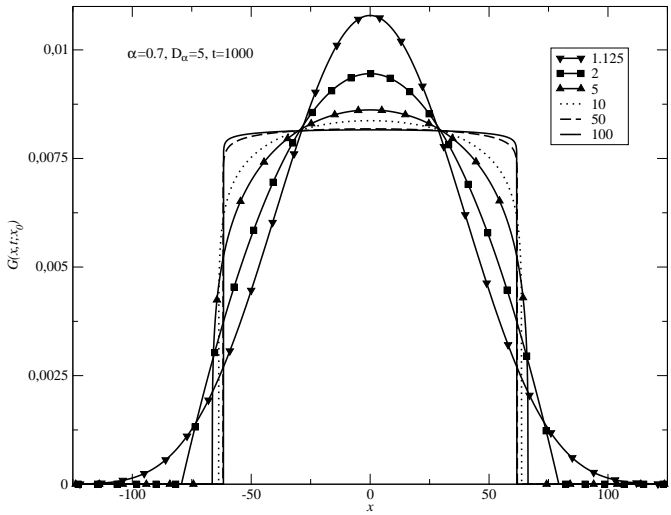}
		\caption{Green's function for SM model in the case of $r>1$ with finite support, $W=154.7$ for $r=1.125$, $79.3$ for $r=2$, $66.4$ for $r=5$, $63.7$ for $r=10$, $61.8$ for $r=50$ and $61.7$ for $r=100$, here $x_0=0$. Values of the parameter $r$ are given in the legend.\label{fig1}}
\end{figure}

The function (\ref{eq12}) has a finite support for $r>1$, so the probability of finding the particle differs from zero only in the interval  
	\begin{displaymath}
x\in\left( x_0-W(t),x_0+W(t)\right)\;,
	\end{displaymath}
where $W(t)=Bt^{\alpha/2}$, $B=\sqrt{\frac{2D_\alpha (3r-1)}{r-1}}$ (see Fig. \ref{fig1}). The boundaries of the interval move with the speed $v_g$ given by the relation
	\begin{equation}\label{eq16}
v_g=\frac{dW(t)}{dt}=\frac{\alpha B}{2t^{1-\alpha/2}}\;.
	\end{equation}

The finiteness of $v_g$ ensures that there is a minimum time $T_{x_2,x_1}$ of the passing of the particle from the point $x_1$ to $x_2$, which is given by the relation
	\begin{equation}\label{eq17}
T_{x_2,x_1}=\left[\frac{(x_2-x_1)^2}{B^2}\right]^{1/\alpha}\;.
	\end{equation}
Using (\ref{eq6}), (\ref{eq8}), (\ref{eq12})  and (\ref{eq17}), we obtain
	\begin{eqnarray}\label{eq18}
\lefteqn{R_{SM,r>1}(t;x_0,x_M)=}\nonumber\\
&=&\Theta(T_{x_M,x_0}-t)+\Theta(t-T_{x_M,x_0})\frac{2}{|z_r|\sqrt{r-1}}\nonumber\\
&&\times\left(\frac{T_{x_M,x_0}}{t}\right)^{\alpha/2}\;{_2F_1}\left[\frac{1}{2},\frac{-1}{r-1};\frac{3}{2};\left(\frac{T_{x_M,x_0}}{t}\right)^\alpha\right]\;,\nonumber\\
&&
	\end{eqnarray}
where	$\Theta$ is the Heaviside function and ${_2F_1}\left[a,b;c;z\right]$ denotes the hypergeometric function,
	\begin{equation}\label{eq19}
{_2F_1}\left[a,b;c;z\right]=\sum_{n=0}^{\infty}\frac{(a)_n(b)_n}{(c)_n}\frac{z^n}{n!}\;,	
	\end{equation}
$(a)_n$ is the Pochhammer symbol, $(a)_n=\Gamma(a+n)/\Gamma(a)$.
From (\ref{eq7}) and (\ref{eq18}) we obtain
	\begin{eqnarray}
\lefteqn{F_{SM,r>1}(t;x_0,x_M)=}\label{a}\\
&=&\Theta(t-T_{x_M,x_0})\frac{\alpha}{|z_r|\sqrt{r-1}}\nonumber\\
&&\times\frac{(T_{x_M,x_0})^{\alpha/2}}{t^{1+\alpha/2}}\left[1-\left(\frac{T_{x_M,x_0}}{t}\right)^{\alpha}\right]^{\frac{1}{r-1}}\;.\nonumber
	\end{eqnarray}

\subsubsection{The case of $1/3<r<1$}\label{sec.3.1.2}

\begin{figure}
	\centering
	\includegraphics[angle=-90,scale=0.32]{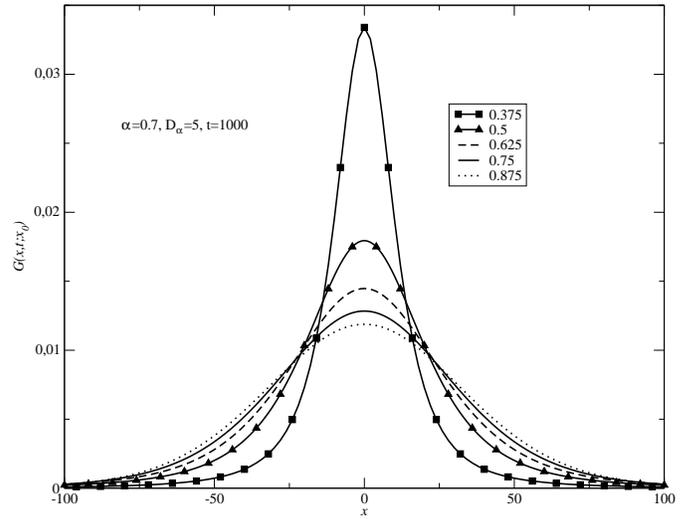}
		\caption{Green's function $G_{SM}$ for $1/3<r<1$. Values of the parameter $r$ are given in the legend.\label{fig2}}
\end{figure}

For this case Green's function is unrestricted. The tails for a specific given $t$ reads $G(x,t;x_0)\sim 1/x^{2/(1-r)}$ when $x\rightarrow\pm\infty$.
The graphs of Greens' functions for this case are presented in Fig. \ref{fig2}.
By  repeating the procedure presented above we obtain
	\begin{eqnarray}\label{eq21}
\lefteqn{R_{SM,r<1}(t;x_0,x_M)=}\nonumber\\
&=&\frac{2}{\sqrt{1-r}|z_r|}\left[\frac{(x_M-x_0)^2}{2D_{\alpha}t^{\alpha}(\frac{3r-1}{1-r})+(x_M-x_0)^2}\right]^{\frac{1}{2}}\nonumber\\
&&\times_{2}F_{1}\left[\frac{1}{2},\frac{1-3r}{2(1-r)};\frac{3}{2};\frac{(x_M-x_0)^2}{2D_{\alpha}t^{\alpha}(\frac{3r-1}{1-r})+(x_M-x_0)^2}\right]\;,\nonumber\\
&&
	\end{eqnarray}
and
	\begin{eqnarray}
\lefteqn{F_{SM,r<1}(t;x_0,x_M)=}\label{b}\\
&=&\frac{\alpha|x_M-x_0|}{|z_r|\sqrt{1-r}}\frac{\left[2D_{\alpha}\left(\frac{3r-1}{1-r}\right)\right]^{\frac{1+r}{2(1-r)}}t^{\frac{\alpha(1+r)}{2(1-r)}-1}}{\left[2D_{\alpha}\left(\frac{3r-1}{1-r}\right)t^{\alpha}+(x_M-x_0)^2\right]^{\frac{1}{1-r}}}\;.\nonumber\\
&&\nonumber
	\end{eqnarray}

Eqs.~(\ref{eq18}),~(\ref{eq19}) and~(\ref{eq21}) provide the same formula for long times, namely for
$t\gg t_{SM}$,
where
$t_{SM}=\left[\frac{(x_M-x_0)^2}{2D_{\alpha}\left|(3r-1)/(1-r)\right|}\right]^{1/\alpha}$,
we obtain
	\begin{equation}\label{eq25}
R_{SM}(t;x_0,x_M)=\frac{\sqrt{2}|x_M-x_0|}{\sqrt{D_{\alpha}(3r-1)}|z_r|}\frac{1}{t^{\alpha/2}}\;,
	\end{equation}
for $r>1/3$ and $r\ne1$.

\subsection{The Tsallis model}

As we mentioned previously, Tsallis Green's function can be treated as a specific case of the Sharma--Mittal one. The results presented in the previous section are valid for the Tsallis model in which
	\begin{equation}\label{eq26}
r=\frac{2}{\alpha}-1\;. 
	\end{equation}
Let us note that the Tsallis model corresponds to the case of $r>1$ for subdiffusion.
As a formality, we present the functions derived in Sec.~\ref{sec.3.1.1}, taking into account Eq.~(\ref{eq26}): 
	\begin{eqnarray}\label{eq27}
\lefteqn{G_{T}(x,t;x_0)=}\nonumber\\
&=&\frac{\sqrt{\alpha}}{\sqrt{2D_\alpha (3-2\alpha)}|z_{2/\alpha-1}|t^{\alpha/2}}\nonumber\\
&&\times\left[\left\{1-\frac{\left(1-\alpha)(x-x_0\right)^2}{(6-4\alpha)D_\alpha t^\alpha}\right\}_{+}\right]^{\frac{\alpha}{2(1-\alpha)}}\;,
	\end{eqnarray}
where
	\begin{displaymath}
z_{2/\alpha-1}=\alpha\frac{\Gamma\left(\frac{\alpha}{2(1-\alpha)}\right)}{\Gamma\left(\frac{1}{2(1-\alpha)}\right)}\sqrt{\frac{\alpha\pi}{2(1-\alpha)}}\;,
	\end{displaymath}
	\begin{eqnarray}\label{eq18t}
\lefteqn{R_{T}(t;x_0,x_M)=}\nonumber\\
&=&\Theta(\tilde{T}_{x_M,x_0}-t)+\Theta(t-\tilde{T}_{x_M,x_0})\frac{2\Gamma\left(\frac{1}{2(1-\alpha)}\right)}{\alpha\sqrt{\pi}\Gamma\left(\frac{\alpha}{2(1-\alpha)}\right)}\nonumber\\
&&\times\left(\frac{\tilde{T}_{x_M,x_0}}{t}\right)^{\alpha/2}\;{_2F_1}\left[\frac{1}{2},\frac{-\alpha}{2(1-\alpha)};\frac{3}{2};\left(\frac{\tilde{T}_{x_M,x_0}}{t}\right)^\alpha\right]\;,\nonumber\\
&&
	\end{eqnarray}
where	
	\begin{displaymath}
\tilde{T}_{x_M,x_0}=\left[\frac{(2-\alpha)(x_M-x_0)^2}{4D_\alpha (3-2\alpha)}\right]^{1/\alpha}\;,
	\end{displaymath}
and the distribution of the FPT is
	\begin{eqnarray}
\lefteqn{F_{T}(t;x_0,x_M)=}\label{c1}\\
&=&\Theta(t-\tilde{T}_{x_M,x_0})\frac{\Gamma\left(\frac{1}{2(1-\alpha)}\right)}{\sqrt{\pi}\Gamma\left(\frac{\alpha}{2(1-\alpha)}\right)}\nonumber\\
&&\times\frac{(\tilde{T}_{x_M,x_0})^{\alpha/2}}{t^{1+\alpha/2}}\left[1-\left(\frac{\tilde{T}_{x_M,x_0}}{t}\right)^{\alpha}\right]^{\frac{1}{r-1}}\;.\nonumber
	\end{eqnarray}	

For long times
$t\gg t_{T}$,
where
$t_{T}=\left[\frac{(2-\alpha)(x_M-x_0)^2}{4D_{\alpha}(3-2\alpha)}\right]^{1/\alpha}$,
Eq. (\ref{eq18t}) can be approximated as follows
	\begin{displaymath}
R_{T}(t;x_0,x_M)=\frac{|x_M-x_0|\sqrt{2(1-\alpha)}\Gamma\left(\frac{1}{2(1-\alpha)}\right)}{\alpha\sqrt{\pi D_{\alpha}(3-2\alpha)}\Gamma\left(\frac{\alpha}{2(1-\alpha)}\right)}\frac{1}{t^{\alpha/2}}\;.
	\end{displaymath}

\subsection{The Gauss model}

Green's function for the Gauss entropy model can be obtained from (\ref{eq12}) in the limit $r\rightarrow 1^-$ and reads
	\begin{equation}\label{eq29}
G_{G}(x,t;x_0)=\frac{1}{2\sqrt{\pi D_\alpha t^\alpha}}\exp\left(-\frac{(x-x_0)^2}{4D_\alpha t^\alpha}\right)\;.
	\end{equation}	
From (\ref{eq8}) and (\ref{eq29}) we obtain
	\begin{equation}\label{eq30}
R_G(t;x_0,x_M)={\rm erf}\left(\frac{x_M-x_0}{\sqrt{4D_{\alpha}t^{\alpha}}}\right)\;,
	\end{equation}
where ${\rm erf}(u)$ is the error function
	\begin{equation}\label{eq31}
{\rm erf}(u)\equiv\frac{2}{\sqrt{\pi}}\int_0^x{\rm e}^{-u^2}du=\frac{2}{\sqrt{\pi}}\sum_{n=0}^\infty\frac{(-1)^nu^{2n+1}}{(2n+1)n!}\;.
	\end{equation}
Using (\ref{eq7}) and (\ref{eq30}) we get
	\begin{equation}\label{d}
F_G(t;x_0,x_M)=\frac{|x_M-x_0|}{2\sqrt{\pi D_{\alpha}}}\frac{1}{t^{1+\alpha/2}}{\textrm e}^{-\frac{(x_M-x_0)^2}{4D_{\alpha}t^{\alpha}}}\;.
	\end{equation}
For $t\gg t_G$, where $t_G=\left[\frac{(x_M-x_0)^2}{4D_{\alpha}}\right]^{1/\alpha}$, from (\ref{eq30}) and (\ref{eq31}) we obtain
	\begin{displaymath}
R_G(t;x_0,x_M)=\frac{|x_M-x_0|}{\sqrt{\pi D_{\alpha}}}\frac{1}{t^{\alpha/2}}\;.	
	\end{displaymath}

\subsection{The fractional model}

The Green's function (\ref{eq34}) provides the relation (\ref{eq1}) if
	\begin{equation}\label{da}
\tilde{D}_\alpha=\Gamma(1+\alpha) D_\alpha\;.
	\end{equation}
Applying Eqs. (\ref{eq6}), (\ref{eq34}) and (\ref{eq8}) we get
	\begin{equation}\label{eq35}
R_F(t;x_0,x_M)=1-f_{-1,\alpha/2}\left(t;\frac{x_M-x_0}{\sqrt{\tilde{D}_\alpha}}\right)\;,
	\end{equation}
and from (\ref{eq35}) and (\ref{eq7}) we obtain
	\begin{equation}\label{eq35a}
F_F(t;x_0,x_M)=f_{0,\alpha/2}\left(t;\frac{x_M-x_0}{\sqrt{\tilde{D}_\alpha}}\right)\;.
	\end{equation}

For $t\gg t_F$, where $t_F=\left[\frac{|x_M-x_0|\Gamma(1-\alpha/2)}{\sqrt{2\tilde{D}_{\alpha}}\Gamma(1-\alpha)}\right]^{2/\alpha}$, Eqs.~(\ref{f}), (\ref{defQT}), and ~(\ref{eq35}) give
	\begin{equation}\label{eq37}
R_F(t;x_0,x_M)=\frac{|x_M-x_0|}{\sqrt{D_{\alpha}\Gamma(1+\alpha)}\Gamma(1-\alpha/2)}\frac{1}{t^{\alpha/2}}\;.
	\end{equation}
We should add here, that the equivalent results to Eqs.~(\ref{eq35}), (\ref{eq35a}) and (\ref{eq37}) were previously obtained by Barkai \cite{barkai}.

\section{Comparision of the models\label{sec4}}

In this section we compare the functions obtained from the different models. Putting the functions (\ref{eq18}), (\ref{eq21}), (\ref{eq18t}), (\ref{eq30}), (\ref{eq35}) into (\ref{eq11}) respectively, we get $\left\langle T\right\rangle =\infty$ for all models.
\begin{figure}
	\centering
	\includegraphics[angle=-90,scale=0.31]{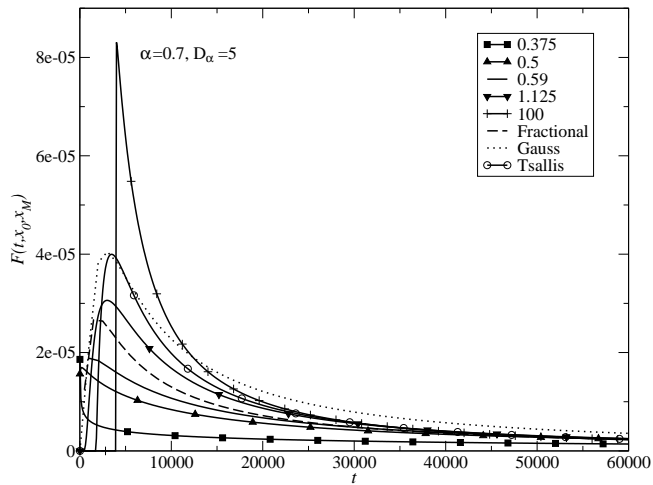}
		\caption{Comparison between $F(t;x_0,x_M)$ for the Sharma--Mittal model for different $r$ values and the other models.\label{fig6}}
\end{figure}
\begin{figure}
	\centering
	\includegraphics[angle=-90,scale=0.322]{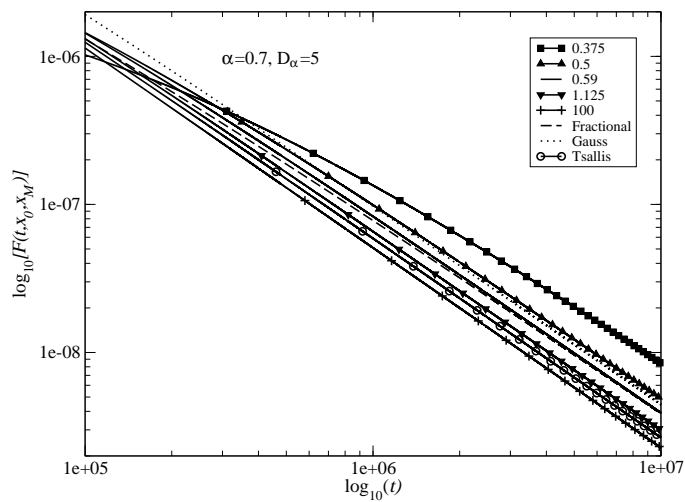}
		\caption{The same situation as in Fig. \ref{fig6} but in the log-log scale for long times.\label{fig7}}
\end{figure}

The probability densities of FPT  are presented in Figs. \ref{fig6} and \ref{fig7}. In both graphs the functions obtained for the fractional model are represented by a dashed line, for the Tsallis model - by a solid one with empty circles and for the Gauss model - by a dotted line; the other lines are assigned to Sharma-Mittal functions with various $r$.
All graphs are prepared for $x_0=0$ and $x_M=100$. The values of the rest of the parameters are given in each figure separately (all quantities are given in the arbitrary chosen units).
In Fig.~\ref{fig6} we have presented graphs for relatively short times calculated for all models studied in this paper, and in Fig.~\ref{fig7} we have presented these functions for long times (in the log--log scale). 
 We can observe that for a very short $t$ the function $F_{SM}$ is larger for a smaller $r$ value (for $r>1$ this function equals zero for $t<T_{x_M,x_0}$).  For intermediate length times we observe the opposite situation, but the Gauss function is the largest now. For very long times, the situation again changes to the opposite and the largest value takes the function for the smallest $r$ value, whereas the Gauss function takes medial values. In the log--log scale (Fig.~\ref{fig7}) we observe that for sufficiently long times, the tails of $F(t;x_0,x_M)$ for all models and different $r$ values are parallel, and according to the functions (\ref{a}), (\ref{b}), (\ref{c1}), (\ref{d}), and (\ref{eq35a}) there is $F\sim 1/t^{1+\alpha/2}$. Let us note that the functions become parallel for a different time, and this time depends on the values of the `reference time' $t_i$, $i=SM,G,T,F$. The $t_{SM}$ depends heavily upon the parameter $r$ and for the cases presented in the graphs there are $t_{SM}=192417$ for $r=0.375$, $19307$ for $r=0.5$, $7650$ for $r=0.59$, $288$ for $r=1.125$ and $3981$ for $r=100$, the other reference times are: $t_T=1767$, $t_G=7173$, and $t_F=794$. We would like to add that the graphs of the functions $F_{SM}$ for various $1.125<r<100$ are located between the functions corresponding to $r=1.125$ and $r=100$ (for graphical clarity these functions are not presented in graphic figures), for $r\geq 100$ all functions are very similar, and in practice they are difficult to distinguish.
\begin{figure}
	\centering
	\includegraphics[angle=-90,scale=0.33]{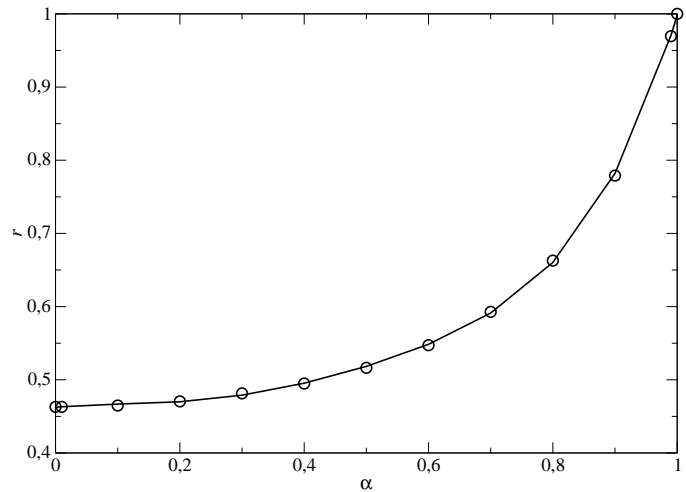}
		\caption{Comparision of the numerical solutions of Eq.~(\ref{eq38}) (circles) and the plot of the function~(\ref{eqr}) (solid line).\label{fig9}}
\end{figure}
Let us note that some of the functions $F$ calculated from the Sharma--Mittal model are very similar to the ones obtained from the fractional equation for sufficiently long times for some $r$ parameters. We are going to find the conditions which ensure that the probability densities of FPT and their cumulative functions will be the same over the long time limit. For $t\gg {\rm max}\{t_{SM},t_F\}$ the agreement condition reads
	\begin{equation}\label{cond}
F_{SM}(t;x_0,x_M)= F_{F}(t;x_0,x_M)\;,
	\end{equation}
which is equivalent to
        \begin{displaymath}
R_{SM}(t;x_0,x_M)= R_{F}(t;x_0,x_M)\;.
	\end{displaymath}
From (\ref{eq25}), (\ref{eq37}) and (\ref{cond}) we get
	\begin{equation}\label{eq38}
\frac{1}{\Gamma(1-\alpha/2)}\frac{1}{\sqrt{\Gamma(1+\alpha)}}=\frac{\sqrt{2}}{\sqrt{3r-1}|z_r|}\;.
	\end{equation}
The numerical solution to Eq. (\ref{eq38}) has a good approximation in the following form (see Fig. \ref{fig9})
	\begin{eqnarray}\label{eqr}
r&=&3.008\alpha^5-5.471\alpha^4+3.768\alpha^3\nonumber\\
&&-0.869\alpha^2+0.101\alpha+0.463\;.
	\end{eqnarray}

\begin{figure}
	\centering
	\includegraphics[angle=-90,scale=0.32]{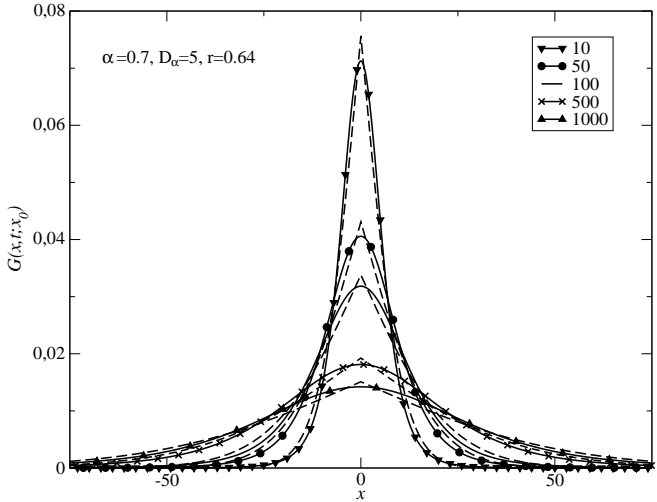}
		\caption{Comparison between Greens' functions for the Sharma--Mittal (solid lines) and the fractional (dashed lines) models for different times given in the legend for $r$ given by (\ref{eqr}).\label{fig3}}
\end{figure}

\begin{figure}
	\centering
	\includegraphics[angle=-90,scale=0.32]{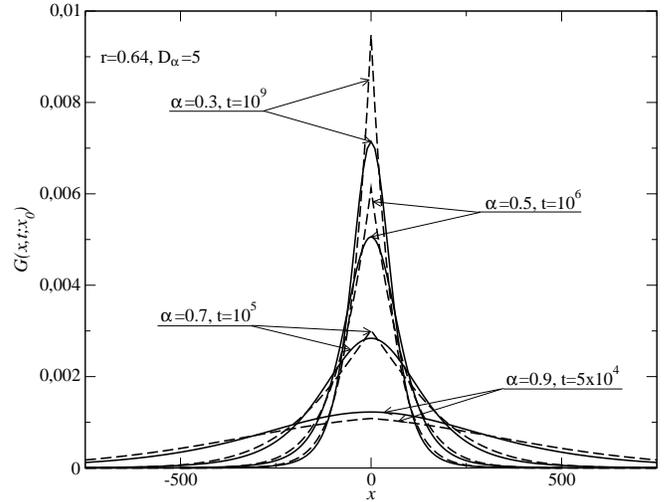}
		\caption{Comparison between Green's function for different $\alpha$ and times given in the legend; $r$ is calculated from (\ref{eqr}) for each $\alpha$ separately. The additional description is the same as in Fig.~\ref{fig3}.\label{fig5}}
\end{figure}

\begin{figure}
	\centering
	\includegraphics[angle=-90,scale=0.32]{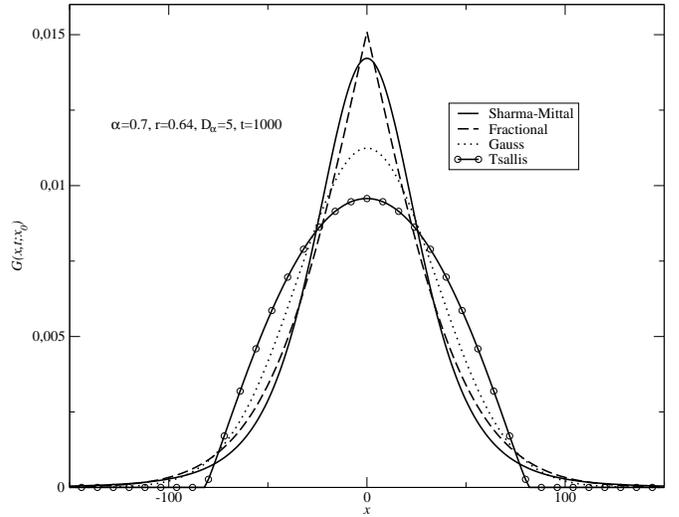}
		\caption{Comparison between the Greens' functions obtained from different models.\label{fig4}}
\end{figure}

We should note that function (\ref{eqr}) gives $r\rightarrow1$ when $\alpha\rightarrow1^-$; this result is to be expected since for $r=\alpha=1$ all models provide Green's function for normal diffusion. 
The functions $G_{SM}(x,t;x_0)$ and $G_{F}(x,t;x_0)$ are also similar to each other for $r$ given by Eq.~(\ref{eqr}), which is shown in Figs.~\ref{fig3} and ~\ref{fig5}. In Fig.~\ref{fig3} Greens' functions for the Sharma--Mittal and the fractional models calculated for $\alpha=0.7$ and various times are presented, in Fig.~\ref{fig5} for different $\alpha$ and $t$.
Obviously, this similarity and the fact that the functions provide the relations (\ref{eq1}) and (\ref{cond}) does not necessarily imply any equivalence between Greens' functions, e.g. for fixed $t$ and for $x\rightarrow\pm\infty$ $G_{SM}(x,t;x_0)\rightarrow(1/|x|)^{2/(1-r)}$, and while the asymptotic form $G_F(x,t;x_0)\sim |x|{\rm exp}(-a|x|^{1/(1-\alpha/2)})$ ($a$ is a positive constant)  has an exponential character and depends on $\alpha$ \cite{mk}.
In Fig.~\ref{fig4} we can observe that the Greens' functions of the Sharma--Mittal and the fractional models are similar for $r$, which are calculated from (\ref{eqr}), in contrast to the Gauss and Tsallis models, which do not fit each other. This is rather obvious, since the subdiffusive Tsallis model corresponds to $r>1$, whereas the Gauss one is only applicable for $r\rightarrow 1^-$. 

\section{\label{f_rem}Final remarks}

We have compared two models describing subdiffusion: the model based on Sharma--Mittal nonadditive entropy, which provides the nonlinear subdiffusion equation with derivatives of a natural order and the one based on Continuous Time Random Walk formalism which provides the linear differential equation with fractional time derivative. Sharma--Mittal formalism contains three parameters $q$, $Q_{SM}$ and $r$ whereas the fractional model is determined by two parameters $\alpha$ and $D_\alpha$.

The main results are as follows:
\begin{enumerate}
	\item\label{pkt1} We have shown that Green's function obtained within Sharma--Mittal entropy formalism fulfils the relation (\ref{eq1}) (this relation is obvious within the fractional model) only if the parameters $q$ and $\alpha$ fulfil the relation (\ref{defq}) and $Q_{SM}$, $D_\alpha$, $\alpha$, $r$ fulfil the relation (\ref{defQi}). We note that relation (\ref{defQi}) is ambigious if the relation between $r$ and the other parameters remains unknown. 
	\item\label{pkt2} We have also shown that the first--passage time probability distribution $F$ and the probability $R$ that a particle has not reached the arbitrary chosen point $x_M$ until time $t$, calculated within the Sharma--Mittal model for $r$ given by the formula (\ref{eqr}) (the other parameters fulfil the relations (\ref{defq}) and (\ref{defQi})), coincide over the long time limit with the ones obtained from the fractional model.  This coincidence is independent of the subdiffusion coefficient $D_\alpha$.
The functions $F$ and $R$ obtained from the Sharma--Mittal model for $r>1$ (which includes the Tsallis model) and for the Gauss entropy model (which corresponds to the limit $r\rightarrow 1^-$ of Sharma-Mittal functions) do not coincide with the results of the fractional model. 
\end{enumerate}

The conditions \ref{pkt1} and \ref{pkt2} have a more general character since a lot of characterisitcs of the system derived within both models will also coincide with each other, as, for example
the time evolution of an amount of substance leaving a semi-infinite medium occupying the region $(-\infty,x_M)$ (then the surface located at $x_M$ can be treated as a fully absorbing membrane) $M(t;x_M)=\int_{-\infty}^{x_M}C(x_0,0)[1-R(t;x_0,x_M)]dx_0$, or the flux $J(t;x_M)=\int_{-\infty}^{x_M}C(x_0,0)F(t;x_0,x_M)dx_0$ flowing through a thin membrane located at $x_M$. 
Let us note that under conditions \ref{pkt1} and \ref{pkt2} the `similarity' of the Greens' functions also occur. This `similarity' is not here a strict definition. It means that the plots of these functions are so similar that we cannot decide in an unambigiuos way which of these fits better the experimental data.
As we mentioned earlier, Green's function can be interpreted as a normalized concentration of a large number of particles starting from the same point at the initial time, thus we can suppose that this `similarity' will occur for concentrations with various initial distributions.

When $r$ is not given by the formula (\ref{eqr}) (which includes the cases $r>1$ and $1/3<r<0.463$), the Greens' functions derived within nonadditive entropy formalism fulfil the relation (\ref{eq1}) and Eqs.~(\ref{defq}), (\ref{defQi}), (\ref{defQT}), (\ref{defQG}) are still valid, but $r$ appears to be independent of $\alpha$ and $D_\alpha$. As we have shown in \cite{kl2012}, subdiffusion described by the Sharma--Mittal entropy formalism has a stochastic interpretation for $1/3<r<1$. Namely, it is a process described by the generalized Langevin equation, in which the strength of the random force is disturbed by the external noise described by the Gamma distribution; then, the parameter $r$ is controlled by the mean--value of the Gamma distribution. When the parameters of both models are related by Eqs. (\ref{defq}), (\ref{defQi}) and (\ref{eqr}), the internal and external noise provides `similar' effects as the noise generating random walk within the fractional model. When $1/3<r<0.463$, the random walk gives the effect that the random walk is more hindered by the external fluctuations than the random walk described by fractional model ($\alpha$ and $D_\alpha$ are the same for both models).
For $r>1$, the support of Green's function (\ref{eq12}) is finite. In consequence, the subdiffusion process is achieved if the velocity of the borders decreases over time according to formula (\ref{eq16}). Let us note that there are two mechanisms of the subdiffusion process. The first one depends on the motion of the borders which limit the area penetrated by the walker, the second is the process of the walker's movement within the allowed region. Since Greens' functions with finite support and nonzero variance cannot describe an infinitely divisable process \cite{feller}, the second mechanism cannot be considered as a random walk process in contrast with the fractional model. For $r\rightarrow\infty$ Green's function approaches a constant distribution \cite{frank5}, thus $r$ can be interpreted as the `measure' of the deviation of the probability density of constant distribution. Therefore, $r$ represents the `uncertainity' of finding the walker in the near--border regions. 

A potential application of the Sharma--Mittal model for $r>1$, is an animal searching for food. A similar problem was considered in \cite{bcmsv}, where the diffusion searching process was assumed to be at two stages. The first stage consists of searching for a diffusive type near the point where the animal stopped its `quick movement' in order to find food, the second stage is the relatively fast movement made by the animal in order to change its searching region. This movement is assumed to be ballistic, e.g. in constant motion. Changing these assumptions, we get the model which --- at least in our opinion --- can be described by the Sharma--Mittal Green's function (\ref{eq12}). Namely, we assume that the animal movement in the restricted region (stage 2) is carried out with a velocity decreasing over time according to formula (\ref{eq16}). This diminishing velocity is connected with the animal's energy, which is gradually lost, especially in the system of a comlpex structure. In stage 1, searching inside the restricted region can be governed by some specific mechanisms, which are not, in fact, random walk. 

Now we shall turn our attention to discussing a method of extracting the subdiffusion parameters occuring in a nonadditive entropy model from experimental data (a similar problem was considered in \cite{ff}).  
The parameters $q$ and $Q_{i}$ can be calculated from Eqs. (\ref{defq}), (\ref{defQi}), (\ref{defQT}) and (\ref{defQG}). The parameters $\alpha$ and $D_\alpha$ are defined by Eq. (\ref{eq1}), but this equation does not lend itself to experimental assingnment. Therefore, one needs to use other functions, which can be measured experimentally. One of such function is $M(t;x_M)$ and $J(t;x_M)$ as defined above. When $r$ is given by (\ref{eqr}), we get $M_{SM}(t;x_M)=M_F(t;x_M)$ and  $J_{SM}(t;x_M)=J_F(t;x_M)$ over a long time limit. Thus, assuming that $0,463<r<1$, each measurement of $M$ and $J$ gives the parameter values  which can be taken from the fractional model. 
The fractional model gives the parameters using a different method, for example, the time evolution of a near-membrane layer \cite{kdm,kdm1}.
To illustrate this we will now calculate the parameters occuring in Sharma--Mittal entropy for the subdiffusion of glucose and sucrose in gel ($1.5$ per cent water solution of agerose). The subdiffusion parameters found in \cite{kdm,kdm1} are: $\alpha=0.90\pm 0.01$, $\tilde{D}_{0.90}=(9.8\pm 1.0)\times 10^{-4} mm^2/s^{0.90}$ for glucose, and $\tilde{D}_{0.90}=(6.3\pm 0.9)\times 10^{-4} mm^2/s^{0.90}$ for sucrose. Taking $r=0.78$ calculated from (\ref{eqr}) and using Eqs.~(\ref{defq}), (\ref{defQi}) and (\ref{da}) we get $q=1.22$ and $Q_{SM}=6.43\times 10^{-4} mm^{2.22}/s$ for glucose, $Q_{SM}=3.93\times 10^{-4} mm^{2.22}/s$ for sucrose (we have omittted here the error calculations).

To extract the subdiffusion parameters from experimental data for the Sharma--Mittal model when $r>1$, it is possible to measure the velocity of the borders $v(t)$, which allows one to determine the parameters $\alpha$ and $B$. We remark that a similar approach was used to find the parameters of the cut-off distribution of postural sway; the estimation of the border locations allows us to calculate the distribution parameters \cite{frank5}. However, in this case, the process was not considered as a stochastic one continuously changing over time, but only two arbitrary chosen `times' (postural sway for the old and young) were taken into account. 

Summarizing the above remarks, the obvious conclusion results from the considerations presented in our paper: subdiffusion cannot be fully characterized by relation (\ref{eq1}) alone. This remark partially agrees with the one presented in \cite{dgn}, where the authors concluded that relation (\ref{eq1}) is not enough in order to designate the process as subdiffusion, but its appropriate stochastic interpretation is also needed.
However, taking into account the fact that there are situations in which the stochastic model has not yet been found, but the models without the stochastic interpretation (such as the ones based on nonadditive entropy) are applicable in describing a process, we assume that both the stochastic as well as the non--stochastic models can be used to describe subdiffusion. Under this assumption it seems to be reasonable to take relation (\ref{eq1}) as the definition of subdiffusion under the conditions that this relation does not provide another important characteristic of the system and its stochastic interpretation may not be obvious. In this context the Sharma--Mittal model appears to be a relatively universal model of subdiffusion where its interpretation and important characteristics (such as first--passage time distribution) depends on the parameter $r$. For special chosen $r$ this model gives very similar  results to the ones provided by the fractional model.

\section*{Acknowledgments}
This paper was partially supported by the Polish National Science Centre under grant No. N N202 19 56 40 (1956/B/H03/2011/40).

\end{document}